\documentclass[12pt]{article}

\usepackage[english]{babel}

\usepackage[letterpaper,top=2cm,bottom=2cm,left=3cm,right=3cm,marginparwidth=1.75cm]{geometry}

\usepackage{amsmath}
\usepackage{graphicx}
\usepackage[colorlinks=true, allcolors=blue]{hyperref}
\usepackage{amssymb}

\title{Power law coupling Higgs-Palatini inflation    with a congruence between physical and geometrical symmetries}
\author{Jos\'e Edgar Madriz Aguilar$^{1}$$^{\dagger}$,   Diego Allan Reyna $^{2}$,  M. Montes $^{1}$}

\begin{document}
\date{}
\maketitle
\begin{center}
    $^{1}$ Departamento de Matem\'{a}ticas, Centro Universitario de Ciencias Exactas e ingenier\'{i}as (CUCEI),
Universidad de Guadalajara (UdG), Av. Revoluci\'on 1500 S.R. 44430, Guadalajara, Jalisco, M\'exico,\\
and\\
$^{2}$ Departamento de F\'{i}sica, Centro Universitario de Ciencias Exactas e ingenier\'{i}as (CUCEI),
Universidad de Guadalajara (UdG), Av. Revoluci\'on 1500 S.R. 44430, Guadalajara, Jalisco, M\'exico. \\
\vspace{0.3cm}

$^{\dagger}$ Corresponding author: José Edgar Madriz Aguilar\\
E-mails:  jose.madriz@academicos.udg.mx, diego.reyna4617@alumnos.udg.mx, mariana.montes@academicos.udg.mx
\end{center}
\begin{abstract}
In this paper we investigate a power law coupling Higgs inflationary model in which the background geometry is determined by the Palatini's variational principle. The geometrical symmetries of the background geometry determine the invariant form of the action of the model and the background geometry resulted is of the Weyl-integrable type. The invariant action results also invariant under the $U(1)$ group, which in general is not compatible with the Weyl group of invariance of the background geometry. However, we found compatibility conditions between the geometrical and physical symmetries of the action in the strong coupling limit. We found that if we start with a non-minimally coupled to gravity action, when we impose the congruence between the both groups of symmetries we end with an invariant action of the scalar-tensor type. We obtain a nearly scale invariant power spectrum for the inflaton fluctuations for certain values of some parameters of the model. Also we obtain va\-lues for the tensor to scalar ratio in agreement with PLANCK and BICEP observational data: $r<0.032$.
\end{abstract}

\\
\vspace{0.3cm}


\section{Introduction}

Non-minimally coupled theories of gravity have been an object of interest in recent years, particularly in the context of cosmological inflation, due to the discovery of the Higgs scalar field \cite{Higgs1, Higgs2}. In the context of the standard model (SM) of particles  the unique scalar field that can play the role of the inflaton by driving inflation is the Higgs scalar field. To achieve a successful inflationary period the so-called non-minimal coupling models of Higgs inflation have been proposed. A non-minimal coupling of the Higgs with gravity of the form $|h|^2R$ is one of the most favored models by observations \cite{Lambda1,Lambda2,Lambda3}. In certain models, these kind of terms could arise from radiative corrections \cite{Lambda3,Lambda4}. Non-minimally coupling Higgs inflationary models aligns excellently with current observations of the cosmic microwave background (CMB) \cite{Lambda1,Lambda2}, and the magnitude of the coupling constant 
 $\epsilon$ between the Higgs field and the Ricci scalar depends of the observed amplitude of perturbations in CMB, resulting $\epsilon \gg 1$ \cite{Lambda5}. One of the problems in particle Higgs models is related with the so called “hierarchy problem”. This consists in the fact that the Higgs boson mass seems to be sensitive to quantum corrections and the bare Higgs mass then needs to be fine-tuned to achieve a physical Higgs boson mass many orders of magnitude smaller than the Planck scale \cite{hierarchy}. 
When the Higgs is coupled to gravity the hierarchy Higgs mass is tackled  through the use of the Palatini variation by several authors \cite{hp1, hp2, hp3, hp4, hp5, hp6, hp7, hp8}. \\

A class of theories that naturally incorporates a scalar field is the  scalar-tensor theories of gravity. Nonetheless, it is not so clear, at least in the Jordan frame, if the scalar field describes gravity or matter \cite{quirozaguilar}. 
 This frame exhibits a non-minimal coupling with the scalar field, however in the Einstein frame a minimal coupling is present \cite{faraoni}. Usually one pass from the Jordan to the Einstein frame by simply implementing a conformal transformation of the metric. The main controversy relies in determining which of the both frames is the physical one \cite{quirozaguilar}. On the other hand, it is a well-known fact that a geometry is characterized by the compatibility condition between the connection and the metric: $\nabla_{\mu} g_{\alpha\beta} = N_{\alpha\beta\mu}$. However, in general the compatibility condition does not remain invariant only under conformal transformations of the metric. Therefore, the usual manner in which we pass from the Jordan to Einstein frame in standard scalar-tensor theories changes the background geometry, and this could be why the physics in one or another frame seems different. Particularly, under conformal transformations a geodesic observer in one frame is not a geodesic observer on the other, and viceversa, see for example \cite{quirozaguilar, faraoni, CR1}. \\ 

This controversy may be alleviated if the background geometry is not fixed apriori as Riemannian. This is the main idea in a recently introduced new kind of scalar-tensor theories known as geometrical scalar-tensor theories of gravity \cite{CR1, CR2}. Geometrical scalar tensor theories of gravity are an approach of scalar-tensor theories of gravity that arise as an attempt to obtain a scalar invariant action under the group of symmetries of the background geometry. Specifically, when a Platini's variational principle is adopted for a scalar-tensor theory the resulting background geometry is non-Riemannian, and as a consequence the group of symmetries that leave invariant the non-metricity condition is bigger than the diffeomorfism group. Thus, the original action is not transforming as a scalar under this extended group. Thus, in this geometrical approach a new action based on the original action is proposed so that it becomes a scalar under the group of symmetries of the background geometry \cite{II1}. In this context, several topics can be investigated in the framework of this new approach, for example, Higgs inflation, the formation of the seeds for cosmic magnetic fields and Dark energy scenarios \cite{II1, II2, II3}. \\

In the case of the majority of non-minimally coupled Higgs inflationary models, it is also used a conformal transformation of the metric to pass to an action free of the non-minimal coupling term, i.e. to the corresponding Einstein frame. This leads to the same controversy of the scalar-tensor theories generated by changing the background geometry, and just like in those cases when the change of the geometry is considered, the original action is no longer an invariant under the new geometrical symmetry group. In this manner,  in this paper we develop a Higgs inflationary model, with a power law non-minimally coupled to gravity term in the action, in the Palatini formalism, where we introduce an  invariant action under the geometrical group of symmetries of the background geometry. In other words, unlike traditional models we use the group of symmetries of the background geometry to obtain the correct action for the model. Thus in  section I we give a brief introduction. Section II is devoted to the construction of the invariant action of the model. In section III we derive the field equations of the particular Higgs inflationary model. In section IV we study the scalar fluctuations of the metric at the end of inflation. Finally, section V is left for some conclusions.
 \\ 

\section{Basic formalism with a non-minimally coupled complex action}

We start with an action for a complex scalar field $\phi$ non-minimally coupled to gravity in the form
\begin{equation}\label{eq1}
{\cal S}=\int d^{4}\sqrt{-g}\Bigl[(1+\epsilon(\phi\phi^{\dag})^{n})M_{pl}^2 R +\frac{1}{2}g^{\alpha\beta}\phi_{,\alpha}\phi^{\dag}_{,\beta}-V(\phi\phi^{\dag}) \Bigr] ,
\end{equation}
where $g$ denotes the determinant of the metric, $R$ is the Ricci scalar curvature, $\epsilon$ is a coupling parameter, $M_{pl}=(16\pi G)^{-1/2}$ is the reduced Planckian mass, the dagger $\dagger$ is denoting transposed complex conjugate, and
\begin{equation}\label{Higgspotential}
    V(\phi\phi^{\dagger})=\frac{\lambda}{2}\left(\phi\phi^{\dagger}-\sigma^2\right)^2
\end{equation}
is the scalar Higgs  potential, with $\lambda=0.129$ and $\sigma=246\, GeV$ being the Higgs vacuum expectation value \cite{III1,III2}. Due to the explicit $U(1)$ global symmetry from the complex scalar field, a gauge derivative $\phi_{:\alpha}$ in terms of a vector field $B_{\mu}$ must be introduced in order to obtain a locally invariant action
\begin{equation}
   {\cal S}= \int d^{4}\sqrt{-g}\Bigl[\left(1+\epsilon(\phi\phi^{\dag})^{n}\right)M_{pl}^2 R +\frac{1}{2}g^{\alpha\beta}\phi_{:\alpha}\phi^{\dag}_{:\beta}-V(\phi\phi^{\dag}) -\frac{1}{4}F_{\mu\nu}F^{\mu\nu}\Bigr] ,\label{eqn:actionsimple3}  
\end{equation}
   where $\phi_{:\mu}=\phi_{,\mu}+i\alpha B_{\mu}\phi$ with $\alpha$ being a gauge parameter of the derivative and  $ F_{\mu\nu}=\partial_{\mu}B_{\nu}-\partial_{\nu}B_{\mu}$ is the  strength field describing the dynamics of the gauge boson field $B_{\mu}$. The action \eqref{eqn:actionsimple3} is invariant under the physical group of symmetries
\begin{eqnarray}\label{rem1}
    \Tilde{\phi}&=&\phi e^{i\gamma \theta(x)},\\
    \label{rem2}
    \tilde{B}_{\lambda}&=&B_{\lambda}-\frac{\gamma}{\alpha}\theta(x)_{,\lambda}\, ,
\end{eqnarray}
where $\gamma$ and $\alpha$ are gauge parameters and the transformation \eqref{rem2} is obtained when $\tilde{\phi}_{:\nu}=e^{i\gamma\theta}\phi_{:\nu}$ is imposed. The expressions \eqref{rem1} and \eqref{rem2} are physical symmetries of the action \eqref{eqn:actionsimple3}. A spontaneously symmetry breaking occurs when $\phi=\phi^{\dagger}$. Thus it is not difficult to verify that after the breaking the minimum of the potential is : $\phi=\pm \sigma$. Hence, it follows that the action \eqref{eqn:actionsimple3} now reads
\begin{equation}\label{rem3}
    {\cal S}=\int d^4x\,\sqrt{-g}\left[(1+\epsilon\phi^{2n})M_{pl}^{2}\,R+\frac{1}{2}g^{\alpha\beta}\phi_{:\alpha}\phi_{:\beta}-V(\phi)-\frac{1}{4}F_{\mu\nu}F^{\mu\nu}\right].
\end{equation}
The Palatini's variational principle, i.e. the variation of the action with respect to the affine connection reveals the background geometry associated to \eqref{rem3} as a non-Riemannian geometry characterized by the compatibility condition: $\nabla_{\lambda}g^{\mu\nu}=-[\ln(1+\epsilon\phi^{2n})]_{,\lambda}\,g^{\mu\nu}$, which after the field rescaling $\varphi=-\ln(1+\epsilon\phi^{2n})$ becomes
\begin{equation}  \nabla_{\lambda}g^{\mu\nu}=\varphi,_{\lambda}g^{\mu\nu}.\label{nonmet}
\end{equation}
Thus the background geometry is  one of the Weyl-Integrable type with null torsion,  where  $\varphi$ playing the role of the Weyl scalar field and $\nabla_{\lambda}$ denotes the Weyl covariant derivative. Notice that there is a geometric symmetry group that leaves invariant (\ref{nonmet}), this is the group of Weyl transformations that reads
\begin{align}
    \Bar{g}&=e^{f}g, \label{eqn:weyltrans1}\\
    \Bar{\varphi}&=\varphi+f, \label{eqn:weyltrans2}
\end{align}
being $f(x^{\gamma})$ a well behaved complex function of the space-time coordinates. \\

The action \eqref{rem3} can be rewritten in terms of the previous field $\varphi$ as
\begin{equation}
   {\cal S}= \int d^{4}\sqrt{-g}\Bigl[ e^{-\varphi}M_{pl}^2\,R +\frac{1}{2}g^{\alpha\beta}\omega(\varphi)D_{\alpha}\varphi D_{\beta}\varphi-U(\varphi) -\frac{1}{4}F_{\alpha\beta}F^{\alpha\beta}\Bigr] ,\label{eqn:rescaleaction}  
\end{equation}
where $U(\varphi)=V[\phi(\varphi)]$, the redefined covariant derivative $D_{\lambda}=\varphi_{,\lambda}+i\alpha A_{\lambda}\varphi$ obeys the relation
\begin{equation}\label{rem4}
    \phi_{:\lambda}=-\frac{e^{-\varphi}(e^{-\varphi}-1)^{\frac{1}{2n}-1}}{\epsilon ^{1/2n}}\,D_{\lambda}\varphi\,,
\end{equation}
where the redefined gauge field is given by
\begin{equation}\label{rem5}
A_{\mu}=\frac{e^{\varphi}}{\varphi}\left(e^{-\varphi}-1\right)B_{\mu},
\end{equation}
and 
\begin{equation}\label{rem6}
    \omega(\varphi)=\frac{e^{-2\varphi}}{\epsilon^{1/n}}\left(e^{-\varphi}-1\right)^{\frac{1-2n}{n}}.
\end{equation}
The strength tensor can also be  written in terms of $A_{\mu}$, and it has the form
\begin{equation}\label{rem7}
    F_{\alpha\beta}=\partial_{\alpha}\left[\left(\frac{e^{-\varphi}}{e^{-\varphi}-1}\right)\varphi A_{\beta}\right]-\partial_{\beta}\left[\left(\frac{e^{-\varphi}}{e^{-\varphi}-1}\right)\varphi A_{\alpha}\right].
\end{equation}
Now, in order to achieve a Weyl-invariance of the action \eqref{eqn:rescaleaction}, we need the Weyl-invariance of the kinetic term in \eqref{eqn:rescaleaction}. Thus, we demand the transformation of the covariant derivative $D_{\lambda}\bar{\varphi}=D_{\lambda}\varphi$. Hence, it follows the rule
\begin{equation}\label{rem8}
    \bar{A}_{\lambda}\bar{\varphi}=A_{\lambda}\varphi+i\alpha^{-1}f_{,\lambda}\,.
\end{equation}
The transformation rule \eqref{rem8} written in terms of $B_{\alpha}$ reads
\begin{equation}\label{rem9}
    \frac{e^{-\bar{\varphi}}-1}{e^{-\bar{\varphi}}}\,\bar{B}_{\lambda}=\frac{e^{-\varphi}-1}{e^{-\varphi}}\,B_{\lambda}+i\alpha^{-1}f_{,\lambda}.
\end{equation}
As the equation \eqref{rem9} comes from demanding a Weyl-invariance then it can be considered as a geometrical symmetry. Now, comparing equations \eqref{rem2} and \eqref{rem9} it is not difficult to note that the physical and geometrical symmetries for the gauge field $B_{\sigma}$ do not match each other. \\

So, in order to the geometric and physical symmetries for the gauge field to be the same we will consider the strong coupling limit, which is obtained when $\epsilon(\phi\phi^{\dagger})^n\gg 1$ in the action \eqref{eqn:actionsimple3}. Thus, we arrive to the condition
\begin{equation}\label{rem10}
\epsilon \gg \frac{1}{(\phi\phi^{\dagger})^n}\gg \frac{1}{\min{\lbrace(\phi\phi^{\dagger})^n\rbrace}}=\frac{1}{\sigma^{2n}}=(0.004)^{2n} .   
\end{equation}
In this limit the Weyl scalar field and the redefined gauge field can be approximated by 
\begin{eqnarray}\label{rem11}
   \varphi =-\ln (\epsilon \phi^{2n}), \\
  \label{rem12}
  A_{\mu}=-\frac{2n\alpha}{\varphi}\,B_{\mu}.
\end{eqnarray}
In the same manner the expressión \eqref{rem4} reduces to
\begin{equation}\label{rem12a}
   \phi_{:\lambda}=-\frac{e^{-\varphi/2n}}{2n\epsilon^{1/2n}}\,D_{\lambda}\varphi. 
\end{equation}
Hence, the equation \eqref{rem6} in this limit reduces to
\begin{equation}\label{remomega}
    \omega(\varphi)=\frac{e^{-\varphi/n}}{(2n)^2\epsilon^{1/n}}.
\end{equation}
Therefore, using \eqref{rem12} in \eqref{rem8} we obtain
\begin{equation}\label{rem13}
  \bar{B}_{\lambda}  =B_{\lambda}-\frac{i}{2\pi\alpha^2}\,f_{,\lambda}.
\end{equation}
Now, by equationg \eqref{rem13} with \eqref{rem2} we obtain that the geometrical and physical symmetries for $B_{\mu}$ match when 
\begin{equation}\label{rem14}
  f=\frac{2\pi\alpha^2}{i}\left(\frac{\gamma\theta}{\alpha}+\kappa\right),  \end{equation}
  where $\kappa$ is an integration constant. Thus, \eqref{rem14} can be considered as a compatibility relation between geometrical and physical symmetries for $B_{\mu}$.\\

  Getting back to the Weyl-invariance of the action \eqref{eqn:rescaleaction}, even if \eqref{rem8} holds, the potential term (the third term) results not Weyl-invariant due to the fact that the determinant of the metric under \eqref{eqn:weyltrans1} transforms as $\sqrt{-\bar{g}}=e^{2f}\sqrt{-g}$. Thus a complete Weyl invariant action $\Bar{{\cal S}}={\cal S}$ (in all of its terms) can be proposed as
\begin{equation}
   {\cal S}_{inv}= \int d^{4}\sqrt{-g}\Bigl[ e^{-\varphi}M_{pl}^2\,R +\frac{1}{2}g^{\alpha\beta}\omega(\varphi)e^{-\varphi}D_{\alpha}\varphi D_{\beta}\varphi-e^{-2\varphi}U(\varphi) -\frac{1}{4}F_{\alpha\beta}F^{\alpha\beta}\Bigr],\label{eqn:scalartensor} 
\end{equation}
where the aditional conditions 
  \begin{align}
    \Bar{\omega}(\bar{\varphi})&\equiv\omega(\Bar{\varphi}-f)=\omega(\varphi),\\
    \Bar{U}(\bar{\varphi})&\equiv U(\Bar{\varphi}-f)=U(\varphi),
\end{align}
must be valid.

\section{The Riemannian limit}

Weyl transformations define an equivalence class between every set of Weyl manifolds together with their metrics, their scalar fields, as well as their gauge fields \cite{II1,CR1,CR2,II2,II3}. For each election of the function $f(x^{\sigma})$ it corresponds a distinct metric which specifies a distinct member ($M,g,\varphi,B$) of the aforementioned class of equivalence. On the other hand, in a Weyl-integrable geometry the differential line element is not preserved under the Weyl group of transformations i.e. $\Bar{ds}^{2}\neq ds^{2}$. However, the particular choice 
 $f=-\varphi$,  specifies a manifold with some effective metric of the form $h_{\alpha\beta}=e^{-\varphi}g_{\alpha\beta}$, such that 
 $\Bar{h}_{\alpha\beta}=h_{\alpha\beta}$. It is not difficult to verify that this effective metric has a metricity condition of the form
 \begin{equation}
\nabla_{\mu}h_{\alpha\beta}=0,  
\end{equation}
therefore, one of the elements of the equivalence class is indeed a Riemannian space-time. On this Riemannian limit the action (\ref{eqn:scalartensor}) takes the following form
\begin{equation}
    {\cal S}=\int d^{4}\sqrt{-h}\Bigl[ M_{pl}^2\,R +\frac{1}{2}h^{\alpha\beta}\omega(\varphi)D_{\alpha}\varphi D_{\beta}\varphi-U(\varphi) -\frac{1}{4}F_{\alpha\beta}F^{\alpha\beta}\Bigr].\label{eqn:RiemannAction}  
\end{equation}
where in the strong coupling limit the strength tensor \eqref{rem7} is given by
\begin{equation}\label{rem15}
    F_{\mu\nu}=-\frac{1}{2n\alpha}\left[\partial_{\mu}(\varphi A_{\nu})-\partial_{\nu}(\varphi A_{\mu})\right].
\end{equation}
On the other hand,  the cosmological principle does not allow the existence of a vector field on cosmological large scales. Thus, it is convenient to use the gauge freedom associated to $A_{\mu}$. With this in mind,  by using  \eqref{rem2} and \eqref{rem12} we arrive to
\begin{equation}\label{rem16}
    \tilde{\varphi}\tilde{A}_{\lambda}=\varphi A_{\lambda}+\frac{\gamma}{2n\alpha^2}\,\theta_{,\lambda}.
\end{equation}
Therefore, by means of the gauge election
\begin{equation}\label{rem17}
    \theta_{,\lambda}=-\frac{2n\alpha^2}{\gamma}\varphi A_{\lambda},
\end{equation}
it can be verified that $\tilde{\varphi}\tilde{A}_{\lambda}=0$. Thus it follows from \eqref{rem15} that $\tilde{F}_{\alpha\beta}=0$. In this manner, Weyl transforming the action \eqref{eqn:RiemannAction} under the gauge election made, we obtain the effective action 
\begin{equation}
   {\cal S}=\int d^{4}\sqrt{-h}\Bigl[ R +\frac{1}{2}\omega(\varphi)h^{\alpha\beta}\partial_{\alpha}\varphi \partial_{\beta}\varphi-U(\varphi) \Bigr],\label{eqn:RiemannAction2}  
\end{equation}
which is a minimally coupled theory that still is carrying the non-canonical kinetic term. 

\section{A Higgs Inflationary model}
Once that we are placed on the Riemannian limit of our theory, we are in position to formulate a Higgs inflationary model. With his purpose we consider the Higgs potential \eqref{Higgspotential}. As we mentioned before a spontaneous symmetry breaking occurs when  $\phi=\phi^{\dag}$.  Thus, excitations around the ground state of $V(\phi)$ can be written as $\phi(x^{\mu})=\sqrt{\xi}+\zeta(x^{\mu})$ where $\zeta(x^{\mu})$ is the Higgs scalar field associated to $\phi$. 
Analogously, in terms of $\varphi$ we can consider the expansion $\varphi(x^{\nu})=\varphi_v+\mathcal{Q}(x^{\nu})$, with $\varphi_v=-\ln(\epsilon\sigma^{2n})$ being the vacuum value for $\varphi$ and $\mathcal{Q}$ the Higgs field associated to $\varphi$. Employing these expressions we can write $\omega(\varphi)$ and the potential $U(\varphi)$  in terms of $\mathcal{Q}(x^{\mu})$ as
\begin{eqnarray}\label{Ohm}
\omega(Q)&=&\frac{e^{-\varphi_v/n}e^{-\mathcal{Q}/n}}{(2n)^2\epsilon^{1/n}},\\
\label{potu}
      U(Q)&=&\frac{\lambda}{2}\Bigl[\sigma^{2} e^{-Q/n}-\sigma^{2}\Bigr]^{2}.
\end{eqnarray}
Hence, the action \eqref{eqn:RiemannAction2} in terms of the Higgs field $\mathcal{Q}$ is then given by
\begin{equation}
     {\cal S}=\int d^{4}\sqrt{-h}\Bigl[ R +\frac{1}{2}h^{\alpha\beta}\omega(\mathcal{Q})\partial_{\alpha}Q \partial_{\beta}Q-U(Q) \Bigr],\label{eqn:RiemannAction2inQ}
\end{equation}
where using the value of $\varphi_v$ in \eqref{Ohm} the function $\omega(\mathcal{Q})$ reads
\begin{equation}
    \omega(Q)=\frac{\sigma^2e^{-\mathcal{Q}/n}}{(2n)^2}.
\end{equation}
 Now, by means of the field transformation  
\begin{equation}\label{rem18}
  \Phi(x)=\int dQ\sqrt{\omega(Q)},   
 \end{equation}
  we can unitarize the kinetic term in \eqref{eqn:RiemannAction2inQ}. Thus we obtain for the rescaled field $\Phi$ and its corresponding potential 
\begin{equation}
\Phi(x)=-\sigma e^{-Q/2n},    \end{equation}

\begin{equation}
    U(\Phi)=\frac{\lambda\sigma^{4}}{2}\Bigl[\Bigl(\frac{\Phi}{\sigma}\Bigr)^{2}-1\Bigr]^{2}. \label{eqn:Phipot}
\end{equation}
Therefore, the action \eqref{eqn:RiemannAction2inQ} can be written as
\begin{equation}
     {\cal S}=\int d^{4}\sqrt{-h}\Bigl[ R +\frac{1}{2}h^{\alpha\beta}\partial_{\alpha}\Phi \partial_{\beta}\Phi-U(\Phi) \Bigr],\label{eqn:RiemannAction2inPhi}
\end{equation}
where it is important to emphasize that due to the unitarization of the kinetic term the potential energy has been rescaled. This fact makes it possible for the Higgs to have enough energy to inflate the universe.\\

Straight forward calculations show that the action (\ref{eqn:RiemannAction2inPhi}) leads to the field equations
\begin{eqnarray}
    && G_{\alpha\beta}=8\pi G\left[\Phi,_{\alpha}\Phi,_{\beta}+\frac{1}{2}h_{\alpha\beta}(\Phi'^{\mu}\Phi,_{\mu}-2U(\Phi))\right]\label{Geq}\\
    && \Box \Phi+U'(\Phi)=0 \label{eqn:Phieqofm},
\end{eqnarray}
where $U'(\Phi)$ denotes the derivative with respect to $\Phi$. Now we can consider the 3D spatially flat Friedmann-Lemaitre-Robertson-Walker metric
\begin{equation}
    ds^{2}=dt^{2}-a(t)(dx^{2}+dy^{2}+dz^{2}),\label{frwmetric}
\end{equation}
where t $a(t)$ is the cosmological  scale factor. Due to the cosmological principle we can adopt the semi-classical approximation 
\begin{equation}
    \Phi(x,t)=\Phi_{b}(t)+\delta\Phi(x,t)\label{eqn:cosmoprin},
\end{equation}
where $\Phi_{b}$ is the homogeneous classical solution to the field equations and $\delta\Phi$ denotes the quantum fluctuations of $\Phi$. As with any other field, the mean fluctuations are null, $\langle\delta\Phi\rangle=\langle\delta\dot\Phi\rangle=0$.
Here we work in the linearized theory, assuming the perturbations to be small $\delta\Phi(x,t)$ so that higher order perturbations are discarded. It follows from (\ref{eqn:Phieqofm}) and (\ref{frwmetric}) that the classical and quantum parts for the inflaton can be splitted into different regimes
\begin{align}     &\ddot\Phi_{b}+3H_{b}\dot\Phi_{b}+U'(\Phi_{b})=0\label{eqn:classfriedenecon}\\
     &\delta\ddot\Phi+3H_{b}\delta\dot\Phi+\frac{1}{a^{2}}(\nabla^{2}\delta\Phi)+U''(\Phi_{b})\delta\Phi=0\label{quantumper}, 
\end{align}
where $H=\dot a/a$ is the Hubble parameter. On the classical side, besides having the relation (\ref{eqn:classfriedenecon}), we also have the following Friedmann equation
\begin{equation}\label{ad1}
    H_{b}=\frac{8}{3}\pi G(\frac{1}{2}\dot\Phi^{2}_{b}+U(\Phi_{b})),
\end{equation}
we can take the time derivative of the previous equation and solve for $\dot\Phi_{b}$, $\dot\Phi^{2}_{b}=-\dot H_{b}/(4\pi G)$, also observe that since $\dot H_{b}=H'_{b}\dot\Phi_{b}$, then $\dot\Phi_{b}=-H'_{b}/(4\pi G)$. Now, under the slow roll condition, we have $|\dot\Phi_{b}|\ll |U(\Phi_{b})|$, and therefore $\dot\Phi_{b}$ becomes
\begin{equation}\label{dotphi}
    \dot\Phi_{b}=-\sqrt{\frac{\lambda}{3\pi G}}\Phi_{b}.
\end{equation}
By integrating both sides of \eqref{dotphi} we can obtain the explicit form for the background field
\begin{equation}\label{ad2}
    \Phi_{b}=\Phi_{be}e^{-\sqrt{\frac{\lambda}{3\pi G}}(t-t_{e})}
\end{equation}
where $\Phi_{be}=\Phi(t_{e})$ with $t_{e}$ denoting the time when inflation ends. The previous expression for $\Phi_{b}x$ near the end of inflation, i.e. $t\approx t_{e}$, can be approximated by a Taylor expansion as
\begin{equation}
    \Phi_{b}(t)\approx \Phi_{be}\Bigl[1+\frac{\lambda}{3\pi G}(t_{e}-t) \Bigr].\label{phibendofinf}
\end{equation}
From \eqref{ad1} and \eqref{ad2}    the explicit form for $H(\Phi_{b}(t))$ reads
\begin{equation}\label{HP}
   H_{b}(t)=\sqrt{\frac{4\pi G\lambda\sigma^{4}}{3}}\Bigl[\frac{\Phi_{be}^{2}}{\sigma^{2}}e^{-\sqrt{\frac{\lambda}{3\pi G}}(t-t_{e})}-1\Bigr].
\end{equation}
Then by substituting the Hubble parameter in terms of $H(a(t))$ we can solve for the scale factor $a(t)$ right after performing an integration from $a$ to $a_{e}$ and from $t$ to $t_{e}$ in order to obtain
\begin{equation}
    a=a_{e}e^{\pi G\sigma^{2}(1-e^{2\sqrt{\frac{\lambda}{3\pi G}}(t-t_{e})})}e^{\sqrt{\frac{4\pi G\lambda\sigma^{4}}{3}}(t-t_{e})}.
\end{equation}
Thus, near the end of inflation the scale factor can be approximated by
\begin{equation}
    a(t)\approx a_{e}\exp{\sqrt{\frac{4\pi G\lambda\sigma^{4}}{3}}(t-t_{e})},\label{aapprox}
\end{equation}
leaving an approximated constant Hubble parameter at the end of this stage.

\section{ The inflaton fluctuations}

Now, in the semi-classical approach we are able to study the inflaton quantum fluctuations $\delta\Phi$. 
The dynamics of these fluctuations is determined by (\ref{quantumper}). With this in mind, we adopt  the standard canonical quantization procedure, and we impose  the canonical commutation relations at every Cauchy surface. According to the action \eqref{eqn:RiemannAction2inQ} the Lagrangian has the form
\begin{equation}
    L=\sqrt{-h}\Bigr[R +\frac{1}{2}\dot\Phi^{2}+\frac{1}{2a^{2}}(\nabla\Phi)^{2}-U(\Phi)\Bigr].
\end{equation}
The canonical conjugate momentum is given by $\Pi=\frac{\partial L}{\partial \dot\Phi}=\sqrt{h}\dot\Phi$.
Therefore the commutation relations are of the  form
\begin{equation}[\delta\Phi(t,\Vec{r}),\delta\dot\Phi(t,\Vec{r'})]=\frac{i}{\sqrt{-h}}\delta^{(3)}(\Vec{r}-\Vec{r'}). \label{wroskian}
\end{equation}
Here we can introduce an auxiliary field $\delta\chi(t,\Vec{r})$ such that
\begin{equation}\label{auxx}
    \delta\Phi(t,\Vec{r})=\exp\Bigl[-\frac{3}{2}H_{b}(t)dt\Bigr]\delta\chi(t,\Vec{r}).
\end{equation}
 In terms of the previous auxiliary field, the dynamical equation for the fluctuations \eqref{quantumper} becomes
 \begin{equation}\label{uch}
     \delta\ddot\chi+\frac{1}{2a^{2}}\nabla^{2}\delta\chi- \Bigl[\frac{9}{4}H^{2}+\frac{3}{2}\dot H -U''(\Phi_{b})\Bigr]\delta\chi=0.
\end{equation}
Now,  we can consider the Fourier expansion of $\delta\chi$ in terms of the annihilation and  creation operators as 
\begin{equation}
    \delta\chi(t,\Vec{r})=\frac{1}{(2\pi)^{3/2}}\int d^{3}k \Bigl[\hat{a}_{k}e^{i\Vec{k}\cdot\Vec{r}}\xi_{k}(t)+\hat{a}^{\dag}_{k}e^{-i\Vec{k}\cdot\Vec{r}}\xi^{*}_{k}(t)\Bigr].\label{modesintermsofanihi}
\end{equation}
With the help of \eqref{uch} and \eqref{modesintermsofanihi} 
the dynamics  of the modes $\xi_{k}$ is governed by
\begin{equation}\label{uchi1}
\ddot\xi_{k}+\Bigl[\frac{k^{2}}{a^{2}}-\frac{9}{4}H^{2}+\frac{3}{2}\dot H-U''(\Phi_{b})\Bigr]\xi_{k}=0.
\end{equation}
It follows from (\ref{phibendofinf}) that  near the end of inflation $U''(\Phi_{e})$ reads
\begin{equation}
    U''(\Phi_{e})=2\lambda(3\Phi_{e}^{2}-\sigma^{2}).
\end{equation}
Therefore the equation \eqref{uchi1}  near to the end of inflation is given by 
\begin{equation}\label{uchi2}
\ddot\xi_{k}+\Bigl[\frac{k^{2}}{a^{2}}-\frac{9}{4}H_{e}^{2}-2\lambda(3\Phi_{e}^{2}-\sigma^{2})\Bigr]\xi_{k}=0.
\end{equation}
Employing \eqref{auxx} and \eqref{modesintermsofanihi} the normalization condition results
\begin{equation}
    \dot\xi_{k}^{*}\xi_{k}-\dot\xi_{k}\xi_{k}^{*}=\frac{i}{a_{e}^{3}}.
\end{equation}
Hence, adopting the Bunch-Davies vacuum, the normalized solution  of \eqref{uchi2} can be written as
\begin{align}
      \xi_{k}(t)= \frac{1}{2}\sqrt{\frac{\pi}{\alpha a_{e}^{3}}}\cdot\mathcal{H}_{\nu}^{(1)}[z(t)].
\end{align}
where  $\nu=\frac{1}{2\alpha}\sqrt{9H_{e}^{2}+24\lambda\Phi_{e}^{2}-8\lambda\sigma^{2}}$ and $z(t)=\frac{k}{\alpha a_{e}}e^{-\alpha(t_{e}-t)}$, where $\alpha=\sqrt{4\pi G\lambda\sigma^{4}}/\sqrt{3}$. \\

The explicit mode functions in the IR sector or superhorizon sector ($-kt<<1$) are dictated by the Hankel function behavior on this regime, which is given by $\mathcal{H}_{\nu}^{(1)}\approx \frac{i}{\pi}\Gamma(\nu)\Bigl(\frac{z(t)}{2}\Bigr)^{\nu}$. Therefore the modes $\xi_{IR}$ have the  form
\begin{equation}
    \xi_{IR}=\frac{1}{2}\sqrt{\frac{\pi}{\alpha a_{e}^{3}}}\frac{i}{\pi}\Gamma(\nu)\Bigl(\frac{z(t)}{2}\Bigr)^{\nu}.\label{xiIR}
\end{equation}
 The mean square fluctuations of the field $\langle 0|\delta\chi^{2}|0\rangle$ are given by 
\begin{equation}
  \langle\delta\chi^{2}\rangle =\frac{1}{2\pi^{2}}\int^{\epsilon_{kh}}_{0}\frac{dk}{k}(\xi_{k}\xi_{k}^{*}),
\end{equation}
where $\epsilon_{kh}$ is the upper limit for the modes inside the horizon. In terms of our original inflaton field $\delta\Phi$, the mean squared fluctuations are given by
\begin{align}
    \langle\delta\Phi^{2}\rangle=\frac{e^{\frac{3}{2}\int H dt}}{2\pi^{2}}\int^{\epsilon_{kh}}_{0}\frac{dk}{k}k^{3}(\xi_{k}\xi_{k}^{*})_{IR}.
\end{align}
Now, with the help of (\ref{xiIR}) we arrive to 
\begin{equation}
\langle\delta\Phi^{2}\rangle=\Bigl(\frac{a}{a_{e}}\Bigr)^{3-2\nu}\frac{2^{2\nu}}{8\pi^{3}}\frac{\Gamma^{2}(\nu)}{(H_{e}a_{e})^{3}}\frac{1}{(H_{e}a_{e})^{-2\nu}}\int^{\epsilon_{kh}}_{0}\frac{dk}{k}k^{3-2\nu},
\end{equation}
Hence, it follows that the  corresponding power spectrum  reads
\begin{equation}
    \mathcal{P}_{s}(k)=\Bigl(\frac{a}{a_{e}}\Bigr)^{3-2\nu}\frac{2^{2\nu}}{8\pi^{3}}\frac{\Gamma^{2}(\nu)}{(H_{e}a_{e})^{3}}\frac{1}{(H_{e}a_{e})^{-2\nu}}\,k^{3-2\nu}.
\end{equation}
Now we are in position to determine the spectral index and the scalar to tensor ratio associated to the inflaton fluctuations in this model.

 \subsection{The spectral index and the scalar to tensor ratio}
 
\begin{figure}[h]
\centering
\includegraphics[width=0.5\textwidth]{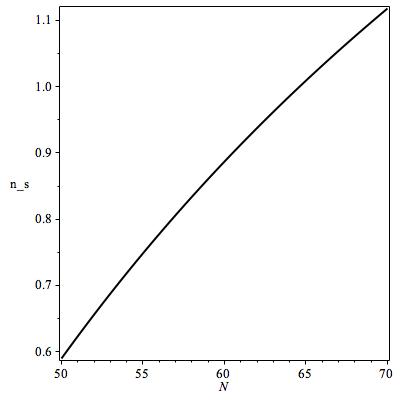}
\caption{This plot shows the behavior of the spectral index $n_s$ as a function of $N$ for  $\epsilon=1.6435\times 10^{-7}$ and $H_e=10^{-5}M_p$. It can be seen that $n$ is increasing respect to $N$ reaching the observational values $n_s\simeq 1$.}
\label{fig0}
\end{figure}

 Once we have obtained the spectral index for inflaton fluctuations we are in position to calculate the scalar to tensor ratio r. In order to do so, we will use the fact without  the strong coupling limit condition, the geometrical scalar fiel $\varphi$ is given by $\varphi=-\ln (1+\epsilon\phi^{2n})$. Thus, considering that $\varphi=\varphi_v+Q$, with $\varphi_v=\ln(1+\epsilon\sigma^{2n})^{-1}$, it is not difficult to verify that with the help of \eqref{Higgspotential} and \eqref{rem6} we arrive to
 \begin{eqnarray}\label{qcut2}
     U(Q)&=&\frac{\lambda}{2}\left[\frac{1}{\epsilon^{1/n}}(1+\epsilon\sigma^{2n})e^{-Q/n}-\sigma^2\right]^2,\\
     \label{qcut3}
     \omega(Q)&=& \frac{(1+\epsilon\sigma^{2n})^2e^{-2Q}}{\epsilon^{1/n}}\left[(1+\epsilon\sigma^{2n})e^{-Q}-1\right]^{\frac{1-2n}{n}}.
 \end{eqnarray}
 Now, using \eqref{rem18} we obtain
 \begin{equation}\label{qcut4}
     \Phi=-\frac{2n}{\epsilon^{1/2n}}\left[(1+\epsilon\sigma^{2n})e^{-Q}-1\right]^{\frac{1}{2n}}
 \end{equation}
 In this manner, the potential \eqref{qcut2} for $n=2$ can be written as
 \begin{equation}\label{qcut5}
     U(\Phi)=\frac{\lambda}{2}\left[\left(\frac{\epsilon\Phi^4+b_0}{b_0\epsilon}\right)^{1/2}-\sigma^2\right]^2,
 \end{equation}
 where $b_0=2^{8}$. The spectral index, is given by
\begin{equation}\label{sin}
n_s=4-\frac{1}{\alpha}\sqrt{9H_e^2+24\lambda\Phi_e^2-8\lambda\sigma^2}.
\end{equation}
In terms of the number of e-foldings 
\begin{equation}\label{qcuw}
N(\Phi)=M_{p}^{-2}\int^{\Phi}_{\Phi_{e}}\frac{U(\Phi)}{U'(\Phi)}\,d\Phi,
 \end{equation}
 where $M_{p}=(8\pi G)^{-1/2}$, the equation \eqref{sin} is given approximately by
\begin{equation}\label{qna1}
    n_s\simeq 4-\frac{1}{\alpha}\sqrt{9H_e^2+\frac{24\lambda\Theta^2}{N}-8\lambda\sigma^2},
\end{equation}
where 
\begin{equation}\label{qna2}
    \Theta=\frac{\sigma\sqrt{2b_0}}{4M_p\epsilon^{1/4}}.
\end{equation}
 Hence, considering a typical value $H_e\simeq 10^{-5} M_p$ and $N=63$ we obtain that for  $\epsilon \in [1.6697\times 10^{-7},1.6435\times 10^{-7}]$ it is achieved the observational bound $n_s=0.968\pm 0.006$ given by PLANCK 2018 cosmological constraints \cite{PRIC}. In the figure [\ref{fig0}] there is a plot showing the increasing behavior of $n_s$ with respect to $N$ for $\epsilon=1.6435\times 10^{-7}$.\\
 
\begin{figure}[h]
\centering
\includegraphics[width=0.7\textwidth]{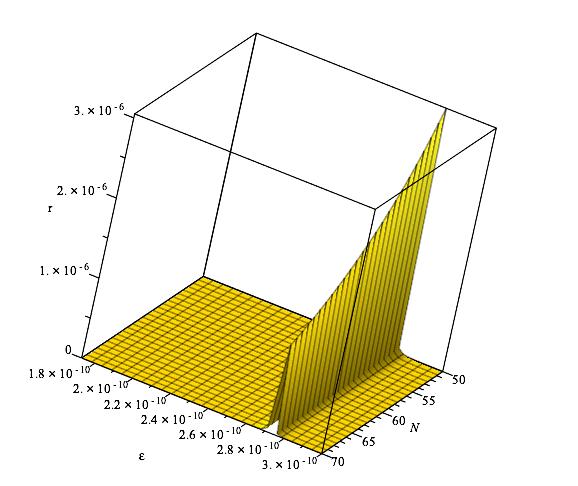}
\caption{This plot shows the behavior of the scalar to tensor ratio $r$ as a function of $N$ and $\epsilon$. It can be seen that $r$ is decreasing respect to $N$ and for values $\epsilon=2.730576184\times 10^{-10}$ the values $r< 0.032$ are reached}
\label{fig1}
\end{figure}

\begin{figure}[h]
\centering
\includegraphics[width=0.5\textwidth]{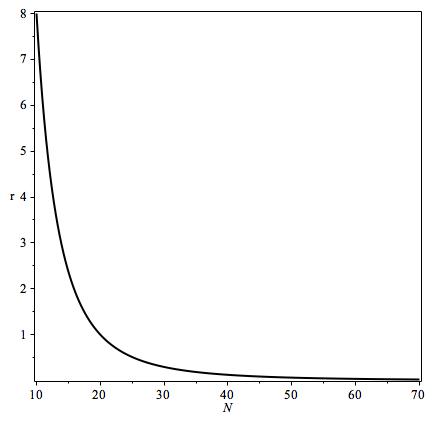}
\caption{This plot shows the relation between the scalar to tensor ratio $r$ and the umber of e-foldings $N$ for $\epsilon=2.730576184\times 10^{-10}$.}
\label{fig2}
\end{figure}

 The scalar to tensor ratio defined by 
 \begin{equation}\label{qcut6}
     r=\frac{16M_{p}^2}{2}\left(\frac{U^{\prime}(\Phi)}{U(\Phi)}\right)^2,
 \end{equation}
 results then
 \begin{equation}\label{qcut7}
     r=\frac{128M_{p}^2\epsilon\Phi^6}{b_0(\epsilon\Phi^4+b_0)\left(\sqrt{\frac{\epsilon\Phi^4+b_0}{\epsilon b_0}}-\sigma^2\right)^2}.
 \end{equation}
Hence, in terms of the number of e-foldings the tensor to scalar ratio can be approximated by
 \begin{equation}
     r\simeq  \frac{64\,\sigma^6}{\sqrt{\epsilon} \,N(4\sigma^4+M_p^{4}N^2)\left(\sqrt{\frac{4\sigma^4}{\epsilon\, M_p^4\,N^2}+1}-\sigma^2\right)^2}
     \end{equation}
According to \cite{Tristan, BPDATA1} observational data corresponding to PLANCK and BICEP the scalar to tensor ratio has the improved limit $r< 0.032$. The value $r=0.032$ is achieved for $N=63$ when $\epsilon=2.730576184\times 10^{-10}$. The same value for r can be obtained for $N=60$ when $\epsilon=2.730573666\times 10^{-10}$. In general, for values of $\epsilon > 2.730576184\times 10^{-10}$ the observational limit $r< 0.032$ is achieved. In the figure [\ref{fig1}] it is shown a plot exhibing  the behavior of the scalar to tensor ratio $r$ as a function of $N$ and $\epsilon$ where it can be seen that $r$ is decreasing respect to $N$ and for values $\epsilon=2.730573666\times 10^{-10}$ the values of $r< 0.032$. The figure [\ref{fig2}] shows the decreacing behavior of $r$ as the e-foldings increase for $\epsilon=2.730576184\times 10^{-10}$.

\section{Final discussion}

In this paper we have studied a Higgs-Palatini inflationary model with a power law coupling. The Higgs scalar field is geometrized by means of the Palatini's principle which in this particular model ensures that the background geometry is a non-Riemannian whose metric compatibility condition can be written in the form of a Weyl-integrable one, where the Weyl geometrical scalar field is the same scalar field that appears in the Higgs potential. This field has a dynamics given by an action that is invariant under the $U(1)$ group of symmetries, which is considered the physical group of symmetries of the action. However, the Weyl-Integrable background geometry also imposes a geometrical group of symmetries: the Weyl group of transformations. The original action is not an scalar under the Weyl group so we propose a new scalar action obeying this requirement but at the same time maintaining the $U(1)$ symmetry. In the new action appears a gauge vector field that allows the action to be an scalar under the Weyl group. Unfortunately, in general, the $U(1)$ requirements on this gauge field are not compatible with the geometrical conditions imposed by the Weyl group. However, in the  strong coupling limit of the Higgs field with gravity appears a compatibility between the geometrical and physical groups of symmetries for the gauge field. This compatibility is one of the main differences of this model with respect other similar ones.\\

Even when we do not start with a scalar-tensor theory of gravity, the new action invariant under the geometrical and physical groups of symmetries can be seen as one of the scalar-tensor gravity type. Thus, in the Riemannian limit of the theory, which corresponds to the Einstein's frame in traditional scalar-tensor theories of gravity, we have calculated the spectrum of inflaton fluctuations, its spectral index and the tensor to scalar ratio. Considering a typical value $H_e\simeq 10^{-5} M_p$ and $N=63$ we have obtained that for  $\epsilon \in [1.6697\times 10^{-7},1.6435\times 10^{-7}]$ it is achieved the observational bound $n_s=0.968\pm 0.006$ given by PLANCK 2018 cosmological constraints \cite{PRIC}. In the figure [\ref{fig0}] the plot shows the increasing behavior of $n_s$ with respect to $N$ for $\epsilon=1.6435\times 10^{-7}$. In  agreement with the observational upper bound provided by using BICEP and PLANCK data: $r<0.032$ \cite{Tristan, BPDATA1}. In our case the value $r=0.032$ is achieved for $N=63$ when $\epsilon=2.730576184\times 10^{-10}$. The same value for r can be obtained for $N=60$ when $\epsilon=2.730573666\times 10^{-10}$. In general, for values of $\epsilon > 2.730576184\times 10^{-10}$ the observational limit $r< 0.032$ is achieved. In the figure [\ref{fig1}] it is plotted the behavior of the scalar to tensor ratio $r$ as a function of $N$ and $\epsilon$. It shows that $r$ is decreasing respect to $N$ and for values $\epsilon=2.730573666\times 10^{-10}$ the values of $r< 0.032$. Also for values of $\epsilon$ ranging in the interval  $[1.6697\times 10^{-7},1.6435\times 10^{-7}]$ the condition $r< 0.032$ is valid, given values of $r$ too small. The figure [\ref{fig2}] shows the decreacing behavior of $r$ as the e-foldings increase for $\epsilon=2.730576184\times 10^{-10}$.

\section*{Acknowledgements}

\noindent J. E. Madriz-Aguilar, Diego Allan Reyna and M. Montes acknowledge SECIHTI
M\'exico and Centro Universitario de Ciencias Exactas e Ingenierias of Guadalajara University for financial support.

\end{document}